# Cloud Provider Capacity Augmentation Through Automated Resource Bartering

Syeda ZarAfshan Goher[a]∗, Peter Bloodsworth[b], Raihan Ur Rasool[c], Richard McClatchey[d]

[a] *NUST School of Electrical Engineering and Computer Science, Islamabad, Pakistan*
[b] *University of Oxford, UK*
[c] *Victoria University, Melbourne, Australia*
[d] *University of the West of England, Bristol, UK*

**Abstract**

Growing interest in Cloud Computing places a heavy workload on cloud providers which is becoming increasingly difficult for them to manage with their primary datacenter infrastructures. Resource limitations can make providers vulnerable to significant reputational damage and it often forces customers to select services from the larger, more established companies, sometimes at a higher price. Funding limitations, however, commonly prevent emerging and even established providers from making continual investment in hardware speculatively assuming a certain level of growth in demand. As an alternative, they may strive to use the current inter-cloud resource sharing platforms which mainly rely on monetary payments and thus putting pressure on already stretched cash flows. To address such issues, we have designed and implemented a new multi-agent based Cloud Resource Bartering System (CRBS) that fosters the management and bartering of pooled resources without requiring costly financial transactions between providers. Agents in CRBS not only strengthen the trading relationship among providers but also enable them to handle surges in demand with their primary setup. Unlike existing systems, CRBS assigns resources by considering resource urgency which comparatively improves customers' satisfaction and the resource utilization rate by more than 50%.The evaluation results provide evidence that our system assists providers to timely acquire the additional resources and to maintain sustainable service delivery. We conclude that the existence of such a system is economically beneficial for cloud providers and enables them to adapt to fluctuating workloads.

## I. Introduction

The ever growing cloud popularity is placing heavy workloads on cloud providers and is making resource provisioning crucialfor them as revealed by recent reports [1]. In the near future, this issue is expected to grow in both number and severity due to the constant expansion of cloud workloads [3, 4]. At present, many providers claim infinite scalability but in practice, this cannot be realized as they all have finite resources in their datacentres. Hardware capacity limitations or periods of maintenance make the satisfaction of dynamically changing levels of requestors' demands difficult to manage. As a result, providers could be forced to reject requests for resources which may damage their reputations for reliability.

In order to scale up providers' capacity, existing cloud solutions (i.e. cloud federation) offer inter-cloud resource sharing on a monetary basis [5, 6, 7]. Price based resource exchange is daunting for providers with limited capital as instead of paying others they may prefer to invest in the expansion of their own data centres. However, in addition to the environmental and economic costs, the scaling of infrastructures has further implications as it may lead towards resource over-provisioning [8].

∗ Corresponding author     E-mail Addresses: zer.afshan@yahoo.com  (S. Goher),

pbloodsworth@ieee.org (P.Bloodsworth),     raihan.rasool@live.vu.edu.au(R.Rasool), richard.mcclatchey@uwe.ac.uk (R.McClatchey)

In response to these problems, it is clear that a flexible resource bartering system is required, which can expand providers' capacities without any need for financial transaction. Resource bartering can help providers to initially borrow resources from their counterparts and then to pay the debt by contributing their resources in future when they become available. This enables providers with tight budgets to augment their physical resources and to thereby handle peak demand from their customers. To realize this concept, an alliance of cloud providers each having different levels of resource demand is created to barter resources with each other. This will not only reduce resource under/over-provisioning problems but will also help emerging providers to ensure long-term service availability. Despite having several benefits, the capability of bartering to assist cloud providers in managing their resource levels has not been well explored yet. Although it is considered an oldermodel of exchange, however, its trend has never gone out of fashion [9]. To demonstrate the validity of bartering in addressing the resource limitation problem of cloud providers, a new multi-agent based Cloud Resource Bartering System (CRBS) for Infrastructure as a Service (IAAS) cloud providers is presented in this paper. It is different from the existing cloud federation in several aspects, for example, it enables a price-free computational resource (i.e. virtual machine)exchange among providers while preserves the autonomy of their technological and business management decisions. To better utilize idle resources, unlike existing systems [5, 6, 7], CRBS prioritizes the most urgent requests and thus enables needy providers to maintain consistent service delivery by getting resources immediately.

The other contributions of this work are: 1) Exploiting multi-agent system's efficiency to gather resources from the participating providers and to provide automatic resource matching. In contrast, current systems force requestors to search and compare each and every resource offer listed in the information directory manually. This is a challenging task when a number of options are available and especially when resources are quickly running out.2) Reserving privileges for altruistic participants (i.e. those who frequently share their resources) within the system. The reward mechanism is necessary to keep the system functional as it motivates the other participants of the system to collaborate actively. To the best of our knowledge, these features altogether have not been incorporated in any of the existing cloud systems and produce an approach that has never been used before. The effectiveness of the CRBS has been assessed by comparing it with the existing inter-cloud resource sharing systems. For evaluation purposes, several testing metrics which are commonly considered crucial for the success of a cloud marketplace have been evaluated. The experimental results validate the devised system ability to effectively barter resources under various scenarios. The remainder of the paper is structured as follows: Section 2 presents the work related to the existing resource sharing systems. Based on the literature review, the architecture and working of CRBS are presented in Sections3 and 4. The evaluation results have been compiled in Section 5 followed by the conclusion and future work.

## 2. Related Work

### 2.1 Resource sharing systems

Cloud computing has found applications in many fields including e-business and e-education [10]. To assist the scientific community in performing the resource-intensive tasks, Cloud@Home was introduced [11]. It enabled the sharing of virtual machines that were owned by individual users to perform complex tasks. Voluntarily sharing resources without an incentive mechanism is, however, vulnerable to trust and reliability problems as unexpected resource termination could significantly degrade the computing performance. To address these problems, trust ensuring policies have been introduced into CRBS which require participants to honour a collective Service Level Agreement (SLA).

The emergence of community cloud and social networks provide alternative means of enabling resource exchange among users. To achieve the common business objectives, a community cloud can be created by multiple organizations working jointly [12]. In that scheme, the resource exchange was made by considering social distance i.e. communities with close acquaintanceship have a low social distance and were solely granted resources. This impedes resource exchange for new communities and weakens their trading relationships with the participants of the other communities. To foster collaboration among exchanging partners, Haiying Shen and Guoxin Liu introduced Harmony [13] which provided a price-assisted reputation mechanism. To avoid being overloaded, well-reputed participants in [13] offer their resources at a higher price which naturally reduces the number of incoming requests. The main objective of limiting resource demand was to maintain providers' reputations by reducing the request rejection rate. However, offering resources at higher prices can be resented by participants and may dissuade them from continually using the same provider.

Moreover, to maintain service delivery besides facing highly variable environmental factors cloud federation was introduced [14]. It creates an illusion of an infinite pool of resources by combining unused capacities of participating providers. The pooled resources in a federated marketplace are traded either by static or dynamic pricing mechanism which limits the entry of emerging cloud providers into the cloud marketplace and could be detrimental to them. Moreover, in the case of a heavy workload, the federation places and serves resource requests on a First Come First Served (FCFS) basis [7]. This brings no benefits for altruistic users (i.e. those who actively share their resources with others) and leaves fewer resource choices for them. Similarly, CometCloud [] was another approach to federate clouds for monetizing spare capacity. Each cloud in this approach has a site manager which outsources incoming tasks as they exceed local capacity of a provider. Federation in such a scheme is created at runtime and each site can join and leave coalition sporadically which may undermine its effectiveness as participants may leave federation at some critical time. This will adversely affect the trading relationship among providers and may exacerbate the trust and collaboration problems.

Besides setting a cloud federation, some other methods of inter-cloud resource sharing also exists such as Super Peer Overlay based Collaborative Mechanism (SPCM) [], and Mandi []. In SPCM overloaded enterprise clouds collaborate with each other and reserve external cloud nodes from their counterparts. Each enterprise cloud in this scheme has a super-peer which takes the resource sharing decisions and delegate resource control. However, no method for resource selection, trust management and resource usage calculation was defined in []. Mandi, on the other hand, sets a marketplace to provide a holistic view of idle resources gathered from multiple providers. Providers in this scheme advertise their free resources and accept bids from customers who can either be cloud providers or general users. The resource sharing in this scheme, however, employed pricing model which is different from CRBS perspective. SC-Share was another framework to help small-scale providers to deal with the excessive resource demands [].To increase individual capacity in SC-Share, when small clouds (SCs) suffer resource outage then instead of buying resources from other public cloud providers they assist each other. Similar to Mandi, this scheme also relies on monetary payments which put pressure on already stretched cash flows and makes additional resource procurement difficult. Moreover, each SC processes incoming requests on FCFS basis and queued new VM requests giving no preferences to the cooperative SCs.

To promote cooperative resource sharing in a cloud environment, F2C was proposed [] that utilizes a group-buying mechanism among cloud users. In this scheme, cloud users form a group to get resources on a group-discounted price. Resource sharing through this strategy requires some users to act as sales agents for a particular provider and to convince other less-knowledgeable users in their group in order to reduce resource prices. F2C promotes Reciprocal Resource Fairness (RRF) strategy to rewardusers in the group according to their contribution level. This scheme is, however, different from CRBS which employs resource bartering instead of group-buying mechanism and promotes resource bartering among IAAS providers rather than general cloud users.

In recent years, the cooperative game theoretic approach has also been employed to exchange cloud services among service providers [15]. Resource exchange in this scheme requires participants to share their profits with all the other participants who involved in constituting a central resource pool. Furthermore, participants are also required to pay an additional cost that is involved in setting the resource pool. This could be undesirable for the majority of participants and may discourage them from the further pooling of their resources. To expand the infrastructure without sharing profit, CloudVO was introduced [16]. The concept of cloudVO has been inherited from the Grid [17] where multiple autonomous administrative domains worked jointly in pursuit of global resource exchange. The proposed architecture [16], however, presented only a limited interaction between two clouds which is too general to map to a real cloud marketplace. In addition to the above-mentioned problems, non-pricing resource sharing systems are also subjected to free riding [18] and whitewashing [19] threats. In free-riding, users (i.e. free-riders) strive to get an unfair resource advantage by refusing to reciprocate with their own services whereas in whitewashing free riders (usually known as whitewashers)change their identities and rejoin networks with a new identity to gain the advantages and rights of a newcomer (i.e. clearing debts and avoiding penalties). To eradicate this selfish behaviour of participants, it is necessary to introduce incentive mechanisms that can motivate participants to practice fair resource exchange. The following subsections elaborate the importance of incentive strategies in greater detail.

*2.2 Incentive mechanisms*

The incentive mechanism in terms of resource sharing can be defined as a method of encouraging participants to exhibit fairness during a transaction [20]. The absence of incentivizing policies leaves no motivation for participants to collaborate and incites them to act selfishly. On the basis of our literature review, incentivizing strategies have been classified into three broad categories, outlined below.

*2.2.1 Direct reciprocity schemes:* require participants to exchange their services simultaneously. This scheme is also recognized as a tit-for-tat strategy which imposes a strict type of bartering i.e. both participants are required to reciprocate desired services concurrently. Participants involved in this mode of exchange receive immediate gains in terms of resources and services [21, 22]. Direct reciprocation addresses the free-riding problem and eliminates the need for a peer to maintain a long trading history. Nevertheless, direct reciprocation is not always advantageous due to its stringency and may often prove impracticable in highly dynamic environments.

*2.2.2 Delayed reciprocity schemes:* create a flexible resource sharing environment by allowing participants to consume resources even when they have no services with which to reciprocate. This scheme considers participants' reputations while allocating resources and is also known as a reputation-based scheme [23, 24]. The reputation shows the trustworthiness of a participant in the resource sharing community and is considered to reward the helpful peers by giving them preference during a transaction. Each node in this scheme rates the other nodes services on the basis of Quality of Service (QoS) and assigns each other a trust value. The trust value is of considerable importance since it is used to distinguish helpful participants from the selfish ones. To gain a fair share of resources, participants in this scheme collaborate with each other to improve their reputations[25.] eBay and Amazon [26] are two popular systems that work on reputation-based schemes. Besides advantages, one of the possible drawbacks related to reputation-based schemes is the possibility of biased behaviour towards newcomers which remain isolated for having no resource sharing history. In order to resolve this issue, a mechanism is required that can help the newly joined participants to build their reputation. One such mechanism has been introduced in our system which helps newcomers to prove their trustworthiness to the other participants of the coalition.

*2.2.3 Payment schemes:* which are further categorized into two broad categories 1) pricing schemes that work on apay-as-you-gobasis,2) non-pricing or soft-incentive schemes that employ virtual currency instead of real money to trade resources [27, 28].The virtual currency is assigned by a third party on the basis of resource exchange statistics and is in the form of tokens[29]. The literature survey accentuated that the soft-incentive schemes appear to be more practical in collaborative networks where the financial transactions are difficult to execute [30]. To illustrate the effectiveness of soft-incentives, Zhengye Liu et al [31] presented the Networked Asynchronous Bilateral Trading model (NABT) which enabled resource sharing among friends over a social network (i.e. Facebook, MySpace). Similarly, Kyle Chard et al [32] utilized the pre-established friendship relationships among participants over a social network (namely Facebook) to share storage capacity and computational resources. However, this method of exchange only promotes resource sharing among friends which is totally different from the CRBS objectives that attempts to support business-to-business resource exchange.

*2.3 Bartering: An alternative to outsourcing*

Bartering evolved long before the invention of currencies as a method of trading resources. This mode of exchange has never gone out of the fashion because of its exciting features and is still used today in many revolutionized forms [33, 34, 35, 36, 37]. Currently, more than 50,000 international enterprises are bartering resources with each other to improve their customers' experiences [38].Despite the many advantages, there are some limitations inherent in bartering such as the absence of money which makes it difficult to measure the value of traded goods. To resolve this issue, modern bartering strategies use barter credits [39] (virtual currency) to regulate bartering transactions (as explained in the previous subsection). Trading through credits is recognized as indirect bartering where a third party known as the trade association acts as a broker and standardizes the bartering criteria and the value of goods that need to be swapped among the participants [40]. This mode of bartering has proved to be better than the direct bartering as it takes the responsibility of finding a trusted trading partner which may often be quite a complex task. The advanced feature of indirect bartering over the primitive one is one of the main reasons we adopted it in CRBS.

*2.4 Significance of multi-agent technology for operating in highly dynamic environments*

Agent-based technology has been widely used to create advanced systems that operate well in open and distributed environments [41.] Agents are function-specific modular components with built-in pro-activeness to adapt to the changing demands. The decentralized architecture of agents enables them to operate in distributed environments where the centralized systems fail to function. In a Multi-agent system (MAS) several agents collaborate to perform complex tasks that are beyond individual capabilities and knowledge [42]. The agents can work autonomously without any break which makes them suitable to operate in highly dynamic environments such as the cloud. For many years, MAS have been incorporated into the Grid and peer-to-peer networks to perform heterogeneous resource management [43, 44, 45]. Agents in these systems share their work status with each other to avoid a single point of failure. To make cloud solutions more intelligent and adaptive, multi-agent technology has been integrated to the cloud domain [46, 47, 48]. To make resource discovery, brokering, and swapping effective the Cloud Resource Bartering System (CRBS) presented in this work has been implemented by means of a multi-agent technology. The software agents employed in CRBS provide a unified view of the idle resources and match them with the incoming requests. Moreover, these agents are self-organized and can change their goals by responding to the varying environmental conditions. For example, the Bartering Agent (BA) sets the resource price dynamically by considering the current resource demand and urgency. The working of the other agents present in CRBS has been elaborated in section 3.

**Table 1: Resource Sharing Strategies Comparison**

| Resource Sharing Strategies | Type of participants involved | Free-Riding Detection | Whitewashing detection | Incentivizing Strategy | Resource Sharing Method |
|---|---|---|---|---|---|
| **Cloud@Home [11]** | Individual users | No | No | No | Voluntarily |
| **Community Cloud [12]** | Individual users | No | No | Friendship | Pricing, voluntarily |
| **Cloud Federation [5]** | Cloud providers | NA | NA | No | Pricing |
| **Reservoir [6]** | Cloud providers | NA | NA | No | Pricing |
| **InterCloud[7]** | Cloud providers | NA | NA | No | Pricing |
| **CometCloud []** | Cloud providers | NA | NA | No | Pricing |
| **Mandi []** | Cloud providers, general cloud users | NA | NA | No | Pricing |
| **SPCM []** | Enterprise cloud | No | No | No | Not Defined |
| **SC-Share []** | Small-scale cloud providers | No | No | No | Pricing |
| **F2C []** | Individual users | Yes | No | Reciprocal Resource Fairness | Pricing, voluntarily |
| **Harmony [13]** | Cloud providers | No | No | Reputation System, virtual credits | Pricing |
| **Game-Theoretic Approach [15]** | Cloud providers | No | No | Sharing Payoff | Pricing |
| **CloudVO[16]** | Cloud providers | No. | No | No | Not Specified |
| **NABT [31]** | Individual users | No | No | No | Voluntarily |
| **Social Cloud [36]** | Individual users | NA | NA | Virtual Credits | Voluntarily |

*2.5 Discussion*

The review of the recent literature illustrated the existing resource sharing strategies along with their advantages and limitations as summarized in Table 1. By analyzing the table, it can be observed that thus far the economic model that is mainly adopted for inter-cloud resource exchange is the pricing model which requires providers to pay an upfront price for getting additional resources. This may be beyond the capability of less financially strong IAAS providers or new startups and can limit their business opportunity. The literature review also highlighted that the existing voluntarily sharing of cloud resources solely supports resource exchange among individual cloud users [11, 12, 31, 36] which is completely different from CRBS objective. Moreover, these systems are prone to a number of issues among these uncooperative peers (i.e. free-riders, whitewashers);a lack of incentivizing mechanisms and bias towards new entrants being the most common ones (as shown in Table 1). Besides these limitations, resource sharing platforms at present are incapable of allocating resources in accordance to the requestor urgency. The absence of such a feature increases the probability of assigning resources to a less urgent requestor while leaving limited resources for the needy ones. Similarly, to improve resource sharing a few systems employed the cooperative game theoretic approach [15]. Users in this approach make binding agreements with each other which make their actions interdependent. Task interdependence often leads to vulnerability where the selfish behaviour of a single player can adversely affect the overall system welfare. Moreover, this scheme also requires users to share their profit with the other participants of the coalition. Revenue sharing with less participative players could be detrimental for the users and may restrict them from further cooperation. To ensure cooperativeness while eradicating selfishness, the CRBS enables service providers to help each other without restricting them to share their profit. It also provides flexibility to participants in adjusting their resource sharing strategy according to marketing trends.

The above discussion illustrates the need of resource bartering for the performance enhancement of cloud providers. As a solution, a new automated and scalable Cloud Resource Bartering System (CRBS) has been implemented in this work. The CRBS strives to address the above-mentioned issues inherited in inter-cloud resource exchanging systems. The evaluation results verify that the CRBS assists providers in maintaining high resource availability whilst keeping their operational costs to a minimum. The following sections will elaborate the detailed architecture and work of the CRBS in greater detail.

## 3. The Cloud Resource Bartering System

*3.1 Case study*

Before going into the detail of our proposed approach, the following case study is used to explain the main functionality of the CRBS. Consider two providers, CloudA and CloudB as is shown in figure 1. Let us assume that CloudA is struggling with an unusually heavy workload and is only left with a few resources. To handle this situation, CloudA requires additional resources which necessitate a heavy investment in the underlying infrastructure; however, limited capital and immediate need make this option unviable. Capacity issues continue to grow eventually leading to new requests being rejected to protect existing users. This lack of capacity substantially harms the business reputation of the provider and customers go elsewhere. To avoid this outcome, CloudA may borrow resources from some other provider and pay the debt by contributing its own resources in the near future when they become available. Unfortunately, at present, there is no system which can be used by providers to help them to exchange resources without monetary payment. Recently, a few providers have employed dynamic pricing as a substitute for handing over-demand [49]. Dynamic pricing strives to maintain the workload by raising the resource price according to the market conditions [50]. Besides lowering the resource demands, the sporadic increase in price could make customers resentful and jeopardizes customer retention.

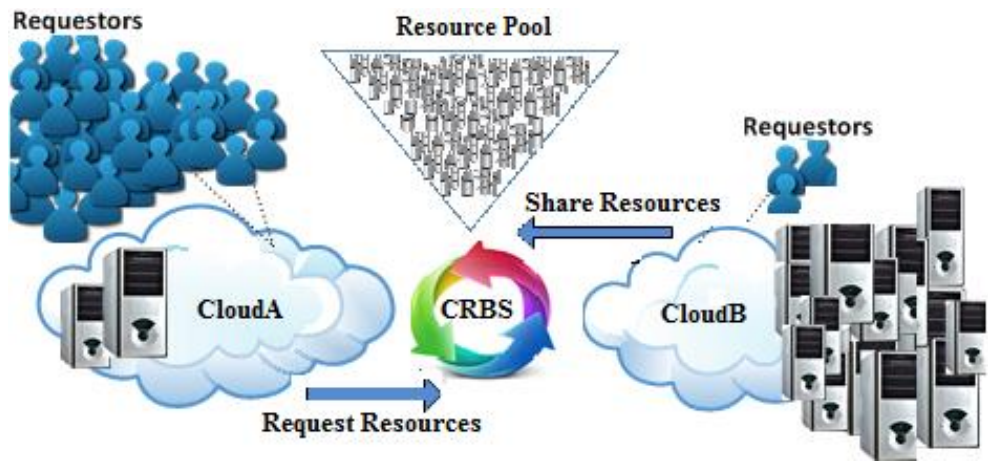

**Fig.1. High-level representation of CRBS**

Now let us consider the scenario from CloudB's perspective, a new startup, which is facing resource underutilization whilst it advertises to build its' user base. During this period, significant hardware capacity may sit idle waiting for demand to grow. When growth occurs, it may be difficult for CloudB to then handle variable demands in parallel with the other challenges that it confronts for being a new entrant to the cloud marketplace. CloudB may, therefore, benefit from a system to which it can delegate dynamic resource management whilst it concentrates on developing its cloud platform.

In response to the aforementioned issues, a new Cloud Resource Bartering System (CRBS) is presented in this work which enables cloud providers to handle unanticipated resource demands by promoting cooperation. It sets up a collaborative platform by registering the participating providers and their contributed resources details. CRBS is deployed by a third party and does not require installation on individual cloud provider premises. Providers with low resource demand (i.e. CloudB) can advertise their free resources details to the CRBS through a web portal to earn barter credits. These credits can further utilize to acquire additional resources during periods of high demands. The contribution of the idle resources to the needy participants enables providers to maintain their service availability by claiming back their resource share, when necessary. Similarly, in the case of high demand, overloaded providers (i.e. CloudA) can make a request using the CRBS interface with an assurance to return the resources as they become available. On receiving resource request, the incorporated bartering mechanism (discussed in the section 3.3.4) starts resource swapping negotiations with the provider on the behalf of the requestor. The negotiation ends either with an SLA establishment to delegate resource control to the requestor or to suggest some other provider for prospective resource gain. Working in this way enables providers such as CloudA and CloudB to seamlessly handle the dynamic expansion and contraction of workloads and thus maintain their credibility. It also promotes trusted relationship amongst participants and creates a "cloud of clouds" that encompasses mutual benefits for providers.

*3.2 System architecture*

In order to efficiently match a resource request to a resource offer, the CRBS is required to perform complex tasks such as resource selection, price negotiation, service level agreement (SLA) management and resource delegation. The successful execution of these tasks requires an intelligent system that can work 24/7 to cope with the changing conditions. One such system known as the multi-agent system (MAS) has been discussed in the literature review along with its main attributes. To utilize the agents' potential, multi-agent technology has been incorporated into our system. There are six main agents comprising CRBS who coordinate each other to promote flexible resource exchange. Agents in CRBS can scale up or down to adapt to fluctuating workloads and are named Platform Agent (PA), User Agent (UA), BlackBoard Agent (BBA), Bartering Agent (BA), Lead Bartering Agent (LBA), and the Transaction Agent (TA) as shown in figure 1. The Platform Agent (PA) performs the administrative tasks such as creation and management of all the other agents present in CRBS. It also registers new users (i.e. cloud providers) and redirects them to a UA which helps to advertise or procure additional resources. On the behalf of requestors, UA negotiates resource exchange with the Bartering Agent (BA) and establishes an SLA.

The Bartering Agent (BA) is one of the main agents of the system and is responsible for controlling the bartering transactions. On the reception of a call-for-proposal (cfp) message, it calculates the transactional price based on the current market situation and negotiates it with the requestor. In the case of a successful transaction, the BA establishes a Service Level Agreement (SLA) between participants and delegates the resource access to the requestor. The BA is managed, launched, and deleted from the system by the Lead Bartering Agent (LBA) as the BA is not required at system startup time. It is required, however, when a new transaction is advertised by a provider. Similarly, the Blackboard Agent (BBA) maintains an architectural model known as the blackboard which displays information about the available resources. The BBA is highly responsive as it continuously updates the blackboard according to new resource advertisements and changing requirements. Likewise, the Transaction Agent (TA) is responsible for keeping auditing records of every transaction in the form of a transactional history. It maintains statistics regarding resource consumption, contribution and the number of barter credits held by each participant. In addition, it is the task of the TA to receive and to update the participants' feedback about the QoS provided. The feedback is used to rank the participants; the significance of ranking will be elaborated shortly.

*3.3 The working of the Resource Bartering Model*

To understand the working of CRBS, it is necessary to analyze its main components which are shown in Figure 2. For the agent-based implementation of our proposed system, the JADE technology was used [51]. Several complex behaviours have been incorporated into agents presented in our model to make them intelligent enough to operate in varying conditions. The CRBS forms a cooperative resource trading group by registering providers through its interface.

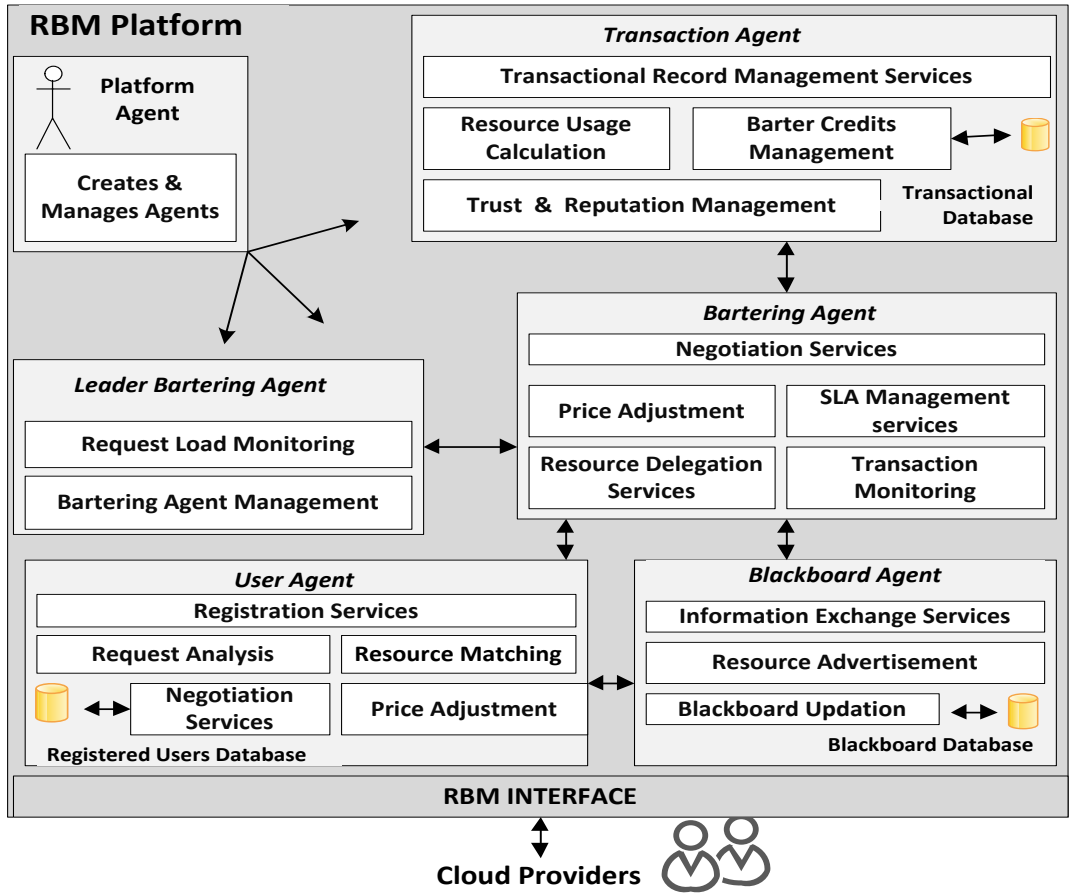

**Fig. 2 CRBS Architecture**

For successful registration, participants accept CRBS terms and conditions and provide their credentials. The users of CRBS are cloud service providers who can act either as a requestor(i.e. requests resources from system) or provider(i.e. share free resources) at a given time. After determining users' status (i.e. requestor, provider), they are forwarded to a UA who helps them to proceed their request. If the user intends to contribute free resources, then the UA takes resource details and executes resource advertisement behaviour whereas in the case of resource acquisition the resource procuring behaviour is exhibited. The behaviours incorporated into UA make it capable of providing multifaceted functionalities and eliminating the need for creating separate agents for each type of user role (i.e. provider, requestor).

### 3.3.1 Resource contribution process

Resource contribution as facilitated by the CRBS enables providers to overcome resource over-provisioning by trading excessive resources (i.e. VMs). Inter-cloud resource exchange is, however, difficult due to the lack of well-defined cloud standards that exacerbates interoperability, vendor lock-in and optimization problems [52]. Therefore, moving an application from one platform to another is quite a complex task. To cope with these issues, a few working standards have been proposed [53, 54.] These standards conceptualize the common attributes of resources and present them semantically. The conceptualized data is then stored in a meta-data repository where it is mapped to the incoming resource requests.

**Table 2: VM Reference Model**

| Specification | Micro Instance | Small Instance | Medium Instance | Large Instance | X-Large Instance |
|---|---|---|---|---|---|
| RAM | 1-2 GB | 4-8 GB | 16-32 GB | 48-64 GB | 80 GB |
| Disk Space | 20-60 GB | 80-240 GB | 320-800 GB | 1-1.5 TB | 2 TB |
| Number of CPUs | 1 | 1-2 | 3-4 | 8-12 | 16 |

Cloud standardization and interoperability is a separate research topic and is consequently beyond the scope of this work. To avoid such an issue, after considering many IAAS providers, a reference model to categorize VMs instances has been created (as is shown in Table 2). The VM classification helps the participating providers to select the type of VM instances for which they want to barter. The Operating System (OS), network bandwidth, and location are not currently considered. These may be considered by some to be important parameters, but they are also prone to the dynamism and this makes the comparison of exchanged resources difficult. It is assumed that providers willing to be the part of CRBS will agree on the given reference model.

**Table 3: VM Instance and Duration Weights**

| Instance Type | Instance Weight ($I_w$) | Sharing Duration | Duration Weight ($D_w$) |
|---|---|---|---|
| Micro | 1 | 1 Week | 1 |
| Small | 2 | 2 Weeks | 2 |
| Medium | 3 | 3 Weeks | 3 |
| Large | 4 | 1 Month | 4 |
| X-Large | 5 | 2 Months | 8 |

The process of resource contribution initiates when the providers select the type of instances that they want to barter from the above reference model. After selection, the instances details are forwarded to UA as an advertisement which includes information such as resource type, the number of required instances, minimum resource price (in terms of barter credits), resource availability location, and resource sharing duration (as is shown in Table 3). The resource price metric included in the resource advertisement is substantial as setting the reasonable price can increase the chances of a successful resource exchange. To guide the provider in determining the suitable amount of barter credits related to a resource offer, a pricing mechanism is employed into the CRBS. It suggests credits/price for a resource advertisement by considering multiple parameters which are instance type, the number of offered instances and the duration for which the resources are shared (as shown in Figure 3). In the absence of a pricing mechanism, it may happen that a provider sets a very high resource price which may keep them underutilized.

To understand the pricing mechanism, consider a scenario in which a provider wants to share ten medium VM instances for the duration of about three weeks. By using the function specified in Figure 3, the calculated instance value would be thirty i.e. (Instance value= Σ (No of Instances * Instance Weight) =10*3= $30) whereas the total barter credits associated with this advertisement would be 90 i.e. (Barter Credits= Instance value * Duration = 30 * 3= 90). After price calculation, the UA suggests it to the provider who could either accept it or reject it by setting resource price independently. The price settlement in this manner will provide greater flexibility to provider owner and would not be detrimental to them as it does not restrict them to share a resource for a fixed price. The UA then appends the

provider's assigned resource price with the original advertisement and forwards it to the Lead Bartering Agent (LBA).

| Barter Credits Calculation Function |
| --- |
| Barter Credits($B_c$) = $I_v * D_w$ |
| where |
| No of Instances = $N_i$ |
| Duration Weight = $D_w$ |
| Instance Weight = $I_w$ |
| Instance value($I_v$) = $\Sigma (N_i * I_w)$ |
| Balance of a participant = $\Sigma B_c$ |

**Fig. 3 Barter Credits Calculation**

After getting a resource advertisement, the LBA creates a bartering agent (BA) for the advertised resource and appends its details to the original resource advertisement and forwards it to the blackboard agent (BBA) to acknowledge resource availability. The appended information helps the requestors to get the details of the bartering agent associated with a resource offer. On getting the resource acknowledgment, the BBA assigns it a unique transaction id and displays it on the blackboard. The transaction id helps to distinguish several offerings presented on the blackboard from each other. The blackboard is one of the important architectural elements of CRBS; it serves as a common information sharing structure for all of the participants. The information displayed on the blackboard helps requestors to select the suitable resources and is comprised of nine attributes which are transaction id, provider details, resource type, the number of available resources, availability location, resource price, sharing duration, provider rank and related BA details to be contacted in case of resource acquisition.

**Table 4: Sample Feedback Parameters**

| No. | Parameters | Rating |
| --- | --- | --- |
| 1. | Availability | |
| 2. | Performance | Excellent (10pts), Very Good (9pts), Good |
| 3. | Response Time | (8 pts), Average (5 pts), and Poor (0 pts) |
| 4. | Fulfillment of SLA | |
| 5. | Elasticity | |

The resource contributor rank as displayed on the blackboard defines the trustworthiness and cooperativeness of a participant, and is significant as in general, everyone wants to share and to take resources from well-reputed providers. The rank of a provider is determined by considering requestor's feedback based on the quality of service (QoS) experienced during a transaction. It is calculated based on metrics such as the resource sharing frequency, availability of resources, performance, response time and the fulfillment of the Service Level Agreement (SLA) as is shown in Table 4.

Participants rate each other's services as excellent, very good, good, average and poor at the end of a transaction. For Excellent 10 points, for Very Good 9 points, for Good 8 points, for Average 5 points, and for Poor 0 points are awarded. To keep the ranking process fair, the average rank value of all the previous transactions associated with a particular provider is calculated and is assigned as rank. Any SLA violation or QoS degradation negatively impacts the rank of a participant and may result in a user's membership being suspended or even canceled. A new provider that joins the system is initially assigned zero points for QoS. The points are increased as it provides quality resources to requestors. If a new provider was given maximum points initially for QoS, then it may be less inclined to provide an excellent service. A low ranking, however, motivates providers to improve their service provision to avail cooperation incentives.

### 3.3.2 Handling resource requests

To maintain service availability besides having limited resources, an overloaded provider can forward resource request to the UA. The request includes the following parameters: VM instance type, required number of VM instances, swapping duration, requestor budget, and preferred region and resource urgency. The preferred region parameter in the request defines the region in which the requestor wants to deploy a VM. This helps the UA to reduce the network latency risk by selecting a provider closest to the requestor's location. Urgency level is an option which indicates how urgently the resource is required by the requestor and is significant for regulating bartering negotiations. The urgency value helps the UA, working on behalf of the requestor, to determine what price should be paid to the provider under different conditions i.e. a high urgency requires to invest higher percentages of budget whereas a low urgency leads the requestor to acquire the most economical resource. By neglecting this urgency metric, it would be difficult to estimate the ideal amount of money that needs to be subsidized for resource exchange. To help participants in specifying resource exigency, the urgency value is categorized into six groups which are: i) Immediate (within 1 hour) ii) within 3 hours iii) within 6 hours iv) within 12 hours v) within 18 hours vi) and within 24 hours (i.e. one day).

| Offering Price determination Function for Requestor |
|---|
| Total time duration required to get resource= $T_t$ |
| Requestor's urgency remaining time= $Rt_r$ |
| Percentage of budget needs to invest=$(20+(80-(80 * Rt_r / T_t)))/100$; |
| Estimated Price= Requestor Budget * Percentage of budget needs to be invested |

**Fig.4 Price Determination Function for Requestor**

The main objective for setting the maximum urgency value of one day is to alleviate the chances of resource over-provisioning; a state in which participants gather resources more than is actually required. Predicting levels of demand for longer time periods (i.e. more than one day) may also increase the risk of resource underutilization which can be a serious problem. Price determination on the basis of urgency enables system capability to adjust the resource price dynamically. As the urgency for resource increases (i.e. time reduces) agents specify an increasing percentage of their overall budgets to procure the desired resource. The lower urgency levels (such as 18 hours, 24 hours) indicate that a requestor does not need the resource desperately and therefore provides agents sufficient time to search for the most affordable offerings. With the passage of the time, the urgency values can be fined tuned by the requestors.

To guide the requestor in determining the resource bidding price, a price determination function has been implemented in the UA as is shown in Figure 4. This function recommends the suitable amount of budget that needs to be invested for a transaction by considering requestor's urgency. Being highly responsive to participants is important therefore the pricing function is implemented using a ticker behaviour that adjusts the resource buying price in accordance to the changing time. The total time ($T_t$) in the pricing function (as shown below) represents the deadline set by the requestor to get resource whereas the $Rt_r$ shows the remaining urgency time of the requestor. Similarly, the estimated price represents the amount of budget suggested by CRBS that the requestor needs to invest in a transaction. After price determination, the UA selects the most economical offer from the blackboard according to the calculated price. The coming section will discuss the working of selection function in greater detail.

### 3.3.3 Resource selection process

In order to select the demanded resources, the UA runs a multi-attribute selection function (as is shown in Figure 5) to pick the appropriate offerings from the blackboard. The function is designed to increase the likelihood of getting the most profitable offerings for the requestor. The selection function considers five parameters for effective resource selection which are: i) Number of required resources ($a1$), ii) Resource swapping duration ($a2$), iii) Budget of the requestor ($a3$) iv) Requestor urgency ($a4$) and v) Provider rank ($a5$). The inclusion of these parameters corroborates the selection of affordable resources from a trustworthy provider in accordance with the requestor urgency. After the selection of suitable offerings, a utility value is calculated for each of them (as is shown in figure 5). To compute the utility value, the resource price as set by the provider and the rank of the provider are considered. These metrics altogether ensure the acquisition of reliable and economical resources.

The offer with the highest utility value represents the most feasible resource selection and is therefore opted for negotiation. To increase the chances of resource exchange, the UA selects the top three offerings having the highest utility score. It then sends a call for a proposal message (cfp) to start negotiation with the BA associated with the highest utility offering. However, if that negotiation fails then the UA moves to the other preselected offers and restarts resource swapping negotiations. If the negotiation for all elected offers remains unsuccessful then the UA revisits the blackboard after a fixed time (i.e. five minutes) to enquire new resource availability. Meanwhile, if a new resource offer appears on the blackboard then the UA is acknowledged to restart resource selection.

Function 1: Resource Selection Function

**Input: a1, a2, a3, a4, a5**
**Output: Selected Offer.**
**For each resource offer on blackboard, the UA checks,**
**IF a4 NOT EQUAL 0**
Then
**IF (Number of required resources>= a1)**
Then
**IF (Availability duration >= a2)**
Then
        IF (Resource Price <= a3)
          Then
Price Benefit x1= a3 - Resource price displayed on blackboard
Provider's Rank x2 = a5
        Utility Function U= (x1+x2)
**END IF**
**END IF**
**END IF**
**END IF**

**Fig.5 Resource Selection Function**

### 3.3.4 Resource bartering negotiations

In order to proceed with the resource exchange, after receiving the call for proposal (cfp) message from the UA, the BA calculates the transactional price on behalf of the provider to start a negotiation. Implementing a negotiation similar to a real world marketplace is difficult because of the distinctive objectives of both buyer and seller. To gain individual benefits participants bargain the resource price with each other that generates different scenarios, for instance, the presence of a large number of unsold products forces the seller to desperately trade all of them even at a price less than or equal to the average product's price and therefore provides more benefits to the buyer; however, if the buyer urgently wants a high demand product then most probably the bargaining will end at a price giving more profit to the seller; on the contrary, if both participants have equal urgencies then the conversation will end at a price approximately equivalent to the average product's price. Consequently, each participant gets an equal chance of earning a profit based on the overall market conditions.

To automate unbiased resource exchange by reflecting the same real-world trading scenarios, a dynamic transactional price determination mechanism has been integrated within the BA which attempts to set fair resource price for prospective resource exchange (as is shown in Figure 6). The price is determined by keeping participants' urgencies in view. The purpose of considering the comparative urgencies when calculating the transactional price is to keep the process fair i.e. giving each participant (requestor/provider) an equal chance of earning a profit. One of the possible drawbacks that could degrade the working of the devised pricing mechanism is the risk of deliberately fabricating false urgencies by the participating providers. By setting fake urgencies, the providers intentionally alter their high exigency into low urgency level to obtain resources at the cheapest price. To mitigate this threat, UA has been implemented in such a manner as to invest lower percentages of the overall budget in the case of low urgency (as discussed in the section 3.3.2). This will automatically reduce the chances of getting desired resources instantly and thus brings no gain for the malicious participant.

**Transactional Price Determination Function for Bartering Agent**

Transactional time ($T_t$) = Expected time limit to acquire resources
$P_{max}$= Maximum resource price set by the provider
$P_{min}$= Minimum resource price at which provider wants to share resource
$Rt_p$ = Provider's remaining urgency time
$Rt_r$= Requestor's remaining urgency time
**Input :** $T_t$, $P_{max}$, $P_{min}$, $Rt_p$, $Rt_r$
**Output:** Transactional Price Determination
**whereas**
Resource price based on provider urgency= $P_1$
Resource price based on requestor urgency=$P_2$
$P_1= (((Rt_p / T_t)*P_{max}) + (P_{min} *(1- Rt_p / T_t)))$
$P_2= (((Rt_r / T_t)*P_{min}) + (P_{max} *(1- Rt_r / T_t)))$
Transactional Price= $(P_1 + P_2)/2$

**Fig.6 Transactional Price Determination Function for Bartering Agent**

After calculating the transactional price by using function in Figure 6, the BA forwards it to the UA involved in the negotiation. The UA compares the quoted price with its current budget as calculated by the function specified in Figure 4. If the quoted price is greater than the requestor's budget then a transaction rejection message is forwarded to BA and the UA moves on to the other possible transactions; otherwise, it sends a confirmation message to the BA and requests resource delegation. Before the actual resource exchange, the confirmed transaction is removed from the blackboard so that it cannot be selected by some other participant; an SLA is established then and the resources are delegated to the requestor. The resource exchanging process finally ends by updating participants' accounts based on resource exchange and recording all the transactional details in a database. The effectiveness of the dynamic transactional price calculation mechanism is evaluated by performing several experiments. The coming section will elaborate the evaluation results of those experiments in greater detail.

## 4. System Evaluation

The effectiveness of the CRBS system can be evaluated in terms of its performance in efficiently managing the bartering of idle resources between providers. Resource swapping in a constantly changing environment such as the cloud is not easy as it faces several challenges. These challenges can be categorized into two major categories where the first category represents the risks specific to the cloud such as unpredictable resource demands, resource underutilization, and resource outages whereas the second category includes risks that are inherited to the resource sharing environments. These include common pitfalls such as the presence of uncooperative participants (i.e. free-riders, whitewashers), the instability of the market equilibrium, lengthy negotiations, poor system responsiveness and bias towards a particular participant. These problems can impede the resource exchange, for instance, free-riding, which is one of the greatest threats related to bartering, incites users to exhibit selfishness and thus creates a lack of trust amongst trading partners.

All of these issues have a significant impact on the cloud marketplace and need to be considered carefully. The validation of CRBS against these issues was quite challenging because at present no such system exists with which it can be compared. The only possible way of testing was to compare some of its common features with the existing inter-cloud resource sharing systems. These systems, however, support price-based resource exchange which is totally different from CRBS perspective. Monetary resource exchange is not susceptible to lengthy negotiations, free-riders, and whitewashers as the resources are shared on a fixed non-negotiable price. Subsequently, there is no need to rank participants or to maintain long resource sharing histories. By considering all these factors, the only features of CRBS that can be compared with the existing inter-cloud systems are the resource selection and allocation. The other possible ways to test the CRBS could be a manual approach which involves asking human participants to barter free resources with each other. The manual approach is, however, incapable of dealing with automated negotiations and does not provide any accounting mechanism. Moreover, it has been experienced that as the number of the participants increases it becomes difficult to manage the system manually. Therefore, the manual approach was not considered for the evaluation. The following subsections will consider the experimental setup and evaluation results in more detail.

4.1 Experimental setup

In order to evaluate CRBS performance while operating under various conditions, different datasets were considered to simulate the real cloudmarketplace where customers make requests to a service provider. To keep the evaluation fair, the datasets were generated randomly and they present the number of participants involved in a transaction along with their requirements. On each dataset several experiments (discussed in the next subsections) were performed to determine CRBS ability to match requestors' requirements to a large number of resources available in the resource pool. Each dataset was comprised of two lists where the first one described providers' specifications such as provider details, advertised resources description, resource price, the rank of the provider, and bartering agent details whereas the second list specified requestors' requirements such as resource description, customers budget, and resource urgency. The datasets were further arranged into three main classes which were small, medium and large. Each small dataset taken in the experiment contains a maximum of twenty-five participants (i.e. providers and requestors) whereas for the medium and large datasets the numbers of participants were 50 and 100 respectively. For each type of input datasets, the number of possible transactions is either the number of providers or requestors, whichever is the smaller. For instance, if the number of providers is 50 and there are 10 requestors then, in this case, the number of possible transactions will be 10. Moreover, in the case of an equal number of participants, the expected transactions would be either equal to the number of providers or requestors.The main objective of dataset classification was to generate test cases that can determine CRBS ability to deal with varying workloads and conditions such as limited resource availability while having high demand, low resource demand while having underutilized resources, resource granting conflicts (i.e. multiple requestors want to consume the same resource), the presence of rigid participants and lengthy futile negotiations. The experiments were designed in such a way to scrutinize CRBS performance under these unpredictable conditions. For comparative analysis, the test cases were firstly executed on CRBS and the results were recorded. The same test cases were then executed by setting an inter-cloud resource trading environment working on the cloud federation principles. The description of each of the conducted experiment along with its results will now be discussed in detail.

### 4.1.1 Resource selection efficiency

The success of a resource sharing platform is largely dependent on its ability to match effectively a given request to a large number of available resources within the given time. To assess CRBS efficiency, various experiments were performed to determine the time it takes to analyze a request and to suggest resources from a large resource pool. For experimentation purposes, initially 100 requestors with different resource requirements were considered and the blackboard was gradually populated with an increasing amount of resource offerings i.e. 500 transactions. This was done to check the responsiveness of the system that how well it can deal with the new offers that appeared on the blackboard.

To perform a comparative analysis, the blackboard was first searched using the existing cloud solutions. A majority of these solutions restrict users to search resources manually which consumes considerable time [55, 56]; however, a few of them (Reservoir [6], InterCloud [7],) support automated resource selection considering the resource price. The provider's reputation is also neglected while selecting the resources in these schemes which may lead to the selection of an unreliable provider who may not deliver the desired QoS. Moreover, unlike the CRBS, these systems do not predict the most favorable provider in the case where multiple providers can satisfy a request. This may lead to selecting a provider who will less likely deliver services. The evaluation results also revealed a delay in resource selection using these systems as is shown in Figure 7. Furthermore, it was also experienced that as the number of entries (i.e. resource offers) on the blackboard increased a delay in the resource selection was observed.

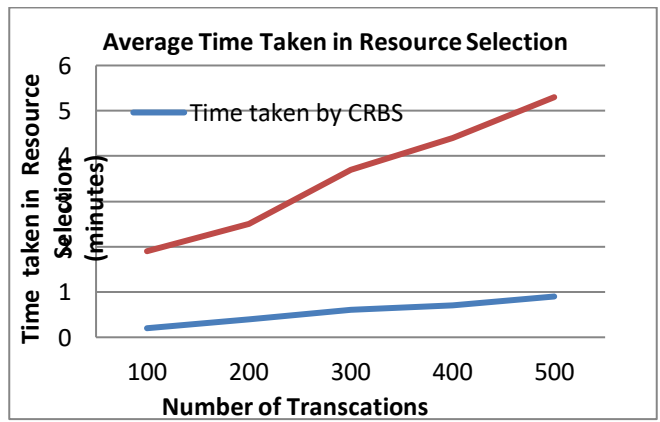

**Fig.7. Average Time Taken in Resource Selection**

However, when the same scenario was executed by CRBS, a reduction in selection time (less than a minute) was experienced even after the addition of new resource offers on the blackboard as is shown in Figure 7. This was due to the efficiency of the resource selection function of CRBS which continuously searches and selects the most suitable resources based on a utility value as discussed in section 3.3.3. From the experimental results, it is evident that the CRBS better assists requestors in selecting the desired resources as compared to the existing inter-cloud systems. Moreover, it was also found that the average time taken in resource selection increases when a relevant transaction is not present on the blackboard. In this case, the UA waits until a new transaction appears which naturally increases the overall resource selection time.

*4.1.2 Handling uncooperative participants*

In a collaborative resource sharing platform, the presence of rational participants who deviate from the standards to maximize their own utility cannot be neglected. The participants that exhibit such selfish behaviour are generally recognized as free-riders or whitewashers as discussed in section 2.1. The existence of unhelpful participants can cause a lack of trust and reduces the overall system utility. However, this problem is only related to systems that promote the voluntary exchange of resources. Such systems are more complex to manage than the existing inter-cloud solutions which grant resources on a monetary basis and therefore do not confront such challenges.

To overcome the risk of free-riding in CRBS, a preventive mechanism was implemented which starts working as soon as a request for resource consumption is received. Before carrying out any transaction, firstly the credibility of the requestor is checked. If the requestor is consistently found in debt, then its request is blocked until the previous debt is cleared. To test the devised mechanism ability to guard against free-riders, a large dataset was taken and a random number of free-riders were injected into the system as is shown in figure8 and 9. The main objective of the testing was to determine the system ability to deal with the selfish participants. The x-axis on the graph shows the number of requestors labeled as participants whereas the y-axis represents the number of confirmed transactions.

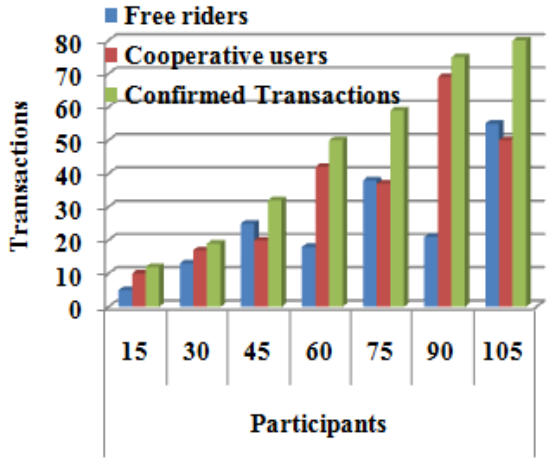
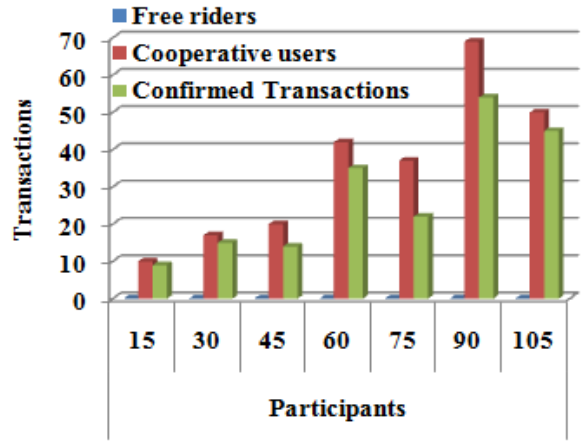

**Fig.8 Number of free riders without preventive mechanism**

**Fig.9 Number of free riders with preventive mechanism**

The experimental results revealed that in the absence of a prevention mechanism, free-riders consume the available resources and thus leave fewer resources for the other cooperative participants (as is shown in Figure 8). On further analysis, it was examined that, when the preventive mechanism was introduced, it did not allow free-riders to consume any resources as is shown in Figure 9. Consequently, due to the blockage of free-riders, the total number of transactions got reduced which is as expected and shows that the mechanism is having the desired impact. Restraining unhelpful participants from executing any transaction reduced market instability as it restricts requestors from accumulating too many resources without reciprocating with their own services.

Additionally, to control whitewashers a reputation-based mechanism was incorporated into the CRBS which ranks participants on the basis of the quality of service that they provided during a transaction (as discussed in section 3.3.1). To deal with whitewashers, whenever a participant joins CRBS it is assigned a low-rank value i.e. zero. In order to gain services, the participants require a higher rank value. Therefore, to increase rank, the newcomers are required to share their resources first. The rank of a participant shows its long-term behaviour and it is not easy to convert a bad reputation into a good one immediately.

Consequently, the inclusion of reputation scheme in CRBS declines the participant's tendency to whitewash.

*4.1.3 Fair resource allocation*

The success of a resource sharing platform is largely dependent on its policy to allocate resources. An effective resource allocation strategy leverages the idle resources and ensures the maximum resource availability in the constituent resource pool. To fairly distribute the resources, CRBS allocates them by considering participant resource sharing history i.e. resources are assigned in proportion to the resources that the participant shared earlier. This promotes fairness as participants get resources according to their record of sharing. In contrast to the CRBS, existing inter-cloud systems (i.e. cloud federation) assign resources on the FCFS basis [7].In these systems, the multiple requests received for a single resource are arranged in a queue which is then served on a First Come First Served Basis. This method of service provisioning can work well with competitive cloud market where the resources are granted on the money. However, it is not suitable for cooperative resource sharing environments where the participants demand incentives for their active participation. The lack of incentivizing mechanism may not leave any motivation for the participants to synergize.

**Table 5: Resource Allocation Scenarios**

| Case No | Rank | Price | Outcome |
|---|---|---|---|
| 1 | Same | Same | Requestor selected randomly |
| 2 | Same | Different | Requestor with highest offering price is selected |
| 3 | Different | Same | Requestor with the highest rank is selected |
| 4 | Different | Different | Requestor with highest offering price is selected |

To incentivize altruism, whenever the resource demand on the CRBS is high and multiple requestors request the same resource simultaneously then the resources are granted to the one having the highest rank. To test the CRBS's resource allocating capability, different test cases were generated. For a better understanding, the categories and outputs of these test cases are summarized in Table 5. The rank column in the table presents the contribution level of requestors who made competing requests for a given resource. The rank value is categorized as same or different; the different rank implies that all the participants who made a request have different ranks and vice versa. The price column shows the number of virtual credits that the requestors were willing to pay for a resource whereas the outcome column specifies the requestor to whom the resource was allocated.

For each category mentioned in Table 5, fifty experiments were performed and the results are summarized as follows: to intelligently handle the request conflicts providers' ranks and offering prices are analyzed. If both parameters are found to be the same, then a requestor is selected randomly (case1). However, if the ranks vary but they offered the same price then to reward the cooperative participant the resource is granted to the requestor having the highest rank (case3). On the other hand, if the requestors offered different prices for the same resource then to maximize the resource owner's profit the transaction is confirmed for the requestor who is willing to pay more (cases2 and case4). A tie can occur when more than one provider is offering the same maximum price. To resolve the tie, the ranks of requestors are examined and the one having the highest rank will be selected to run the transaction whereas the other requestors are forwarded to the other possible transactions. Working in this way, reserves privilege for the helpful participants and persuade the other members of the group to share their resources actively.

*4.1.4 Balancing resource consumption and contribution*

Balancing the type of resources consumed and contributed is necessary for platforms that support the voluntarily sharing of resources. It not only maintains market equilibrium but also ensures the heterogeneous resource availability in the resource pool. The absence of such a mechanism may lead to a situation where a particular type of resource would be excessively present or absent from the market and therefore handling diverse resource demands would be difficult, even if the resource pool is far from empty. The current inter-cloud resource sharing systems, however, do not support this feature as they do not pool resources and solely exchange resources on a monetary basis[6, 7].

The CRBS is designed in such a way that whenever a debtor/participant wants to contribute/offer its resources to clear previous debt then before proceeding with the resource reimbursement process, the current resource demand is checked. If the offered resources are required by the system then they are accepted and the equivalent amount of barter credits is deducted from the debtor's debt. Otherwise, the debtor is restricted to contribute the same indebted resource type; this is done to ensure heterogeneous resource availability in the resource pool. To test the CRBS's ability to maintain resource balance, different experiments have been conducted in which a debtor that owed a debt to a particular type of VM instance tried to clear it by sharing a VM with a different configuration. For instance, the debtor tried to pay a large VM instance debt by contributing micro instances. The evaluation results revealed that under normal conditions (i.e. the resources offered by a debtor was not required by the system) our system did not allow the paying off of a large instance of debt by donating micro instances. Instead, it ensured repayment with the same resource type as that was taken earlier. However, it was also observed that if the debtor offered resources were scarce and was required by the system then the micro instances equivalent to the large instance debt was accepted. This was done to satisfy both participants (i.e. requestor, debtor) and helping them to maintain their resource utilization and service availability rate to a maximum.

*4.1.5 Number of successful transactions*

The number of successful transactions reflects the system's ability to effectively match a request to a resource advertisement. It is a true indicator of the system's capability to successfully make use of idle resources. To judge the ability of the system to dynamically adjust a given workload, several experiments have been conducted on CRBS. The main objective of the experimentation was to calculate the resource utilization and the request satisfaction rate of CRBS under varying workloads. In any condition, the number of successful transactions depends on the resource urgency and the minimum number of participants i.e. requestors/ providers. Based on the number of participants, the experiments were divided into three main categories, which were: i) Experiment 1: the number of providers was greater than number of requestors (i.e. more resources were available than required as is shown in Figure 10), ii) Experiment 2: the number of requestors was greater than number of providers (i.e. fewer resources were available than required as is shown in Figure 11), iii) Experiment 3: equal numbers of both (as is shown in Figure 12).The evaluation results of each of the experiment have been summarized in Table 6.

To simulate an environment similar to the cloud marketplace, requestors with varied resource urgency were considered and the average number of successful transactions for each urgency level was recorded. As discussed previously, the CRBS was designed in such a manner that when the urgency for a resource is high then a higher percentage of the budget is utilized to get resources, thereby thenumber of transactions increased. However, low resource urgency entices to invest a smaller percentage of the overall budget which naturally reduces the number of confirmed transactions.

To perform the first type of experiments, a total of 100 providers and 50 requestors were created. The maximum number of transactions that could be confirmed in this case was equal to the number of requestors. During experimentation, it was observed that due to the low resource demand (i.e. 50 requestors) the resource utilization rate remained low whereas the request satisfaction was much higher as is shown in Figure 10. Moreover, it can be seen from the graph that a large number of transactions were made when the resource urgency was high whereas, in the case of low urgency, the number of transactions got reduced. This verifies the accurate working of urgencies as expected. To perform the second type of experiment, the number of requestors was kept greater than the providers i.e. 100 requestors and just 50 providers. This was done to assess system ability to manage high resource demand while having limited resources.

**Table 6: Resource Utilization and Request Satisfaction Statistics**

| Experimental Statistic | Experiment1 | Experiment2 | Experiment3 |
|---|---|---|---|
| **Number of providers** | 100 | 50 | 100 |
| **Number of requestors** | 50 | 100 | 100 |
| **Number of available resources** | 100 | 50 | 100 |
| **Number of consumed resources in CRBS** | 45 | 42 | 89 |
| **Number of consumed resources in Inter-cloud scenario** | 21 | 30 | 59 |
| **Resource utilization rate of CRBS** | 45% | 84% | 89% |
| **Resource utilization rate in an inter-cloud scenario** | 21% | 60% | 59% |
| **Request satisfaction rate of CRBS** | 90% | 42% | 89% |
| **Request satisfaction rate of an inter-cloud scenario** | 42% | 30% | 59% |
| **Percentage difference of request satisfaction rate** | 73 % | 33% | 41% |
| **Average percentage difference of request satisfaction rate** | 51% | | |

The maximum number of transactions that could be confirmed in this case was equal to 50 (the number of providers). By considering the table 6, it can be observed that the resource utilization rate, in this case, was high i.e. more than 80%. On the other hand, the request satisfaction rate was low due to the limited resource availability as is shown in Figure 11. In the third experiment, an equal number of requestors and providers (i.e. 100) were taken. The resource utilization and the request satisfaction rate in this situation were much higher i.e. about 90%. This was due to the system ability to effectively barter the available resources as is shown in Figure 12.

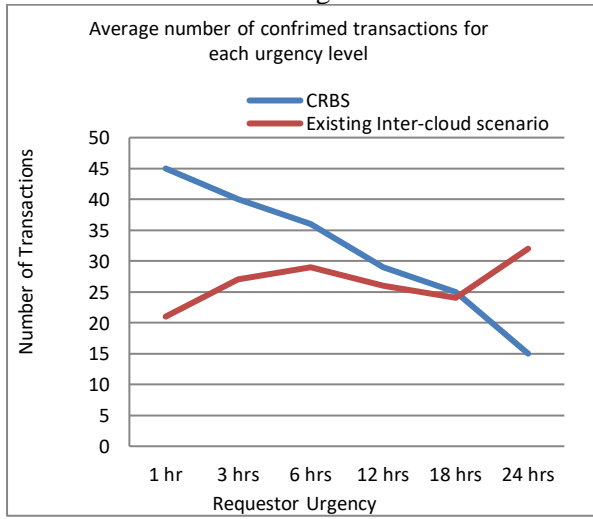

**Fig.10 Request satisfaction rate in case of low resource demand**

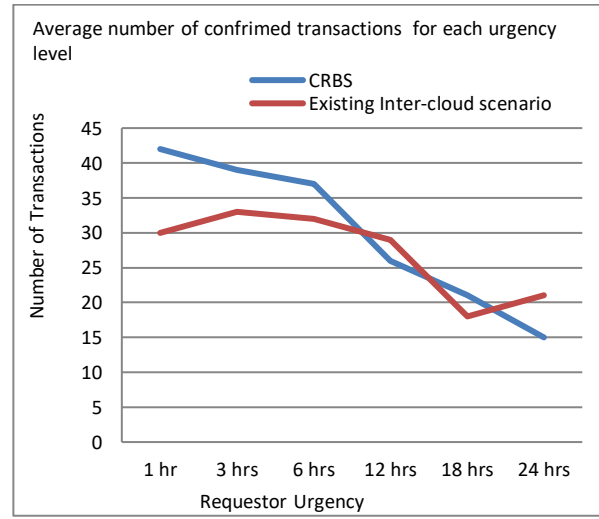

**Fig.11 Request satisfaction rate in case of less resource availability**

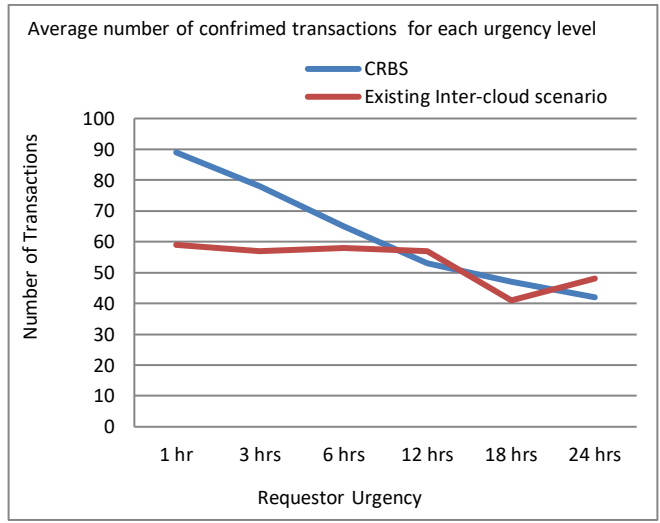

**Fig.12 Request satisfaction rate in case of high resource demand and high resource availability**

To compare CRBS performance with the existing systems, the same experiments were then conducted on the simulated inter-cloud environment where the resources are granted on the FCFS basis. In cases of high demand, requests in such schemes are queued and are satisfied on the basis of requests' arrival time and the offered price. A central entity is known as a cloud exchange which is present in an inter-cloud scenario assigns resources whilst ignoring the requestor reputation and the resource urgency as the resources are not granted voluntarily. To make an accurate comparison possible, the datasets for evaluating both of the systems were kept the same. However, the evaluation results highlighted a reduced number of transactions in that setup which was probably due to the resource exigency being ignored during resource allocation. Resource assignment on the FCFS basis leaves fewer resources for the more urgent requestors and requires them to wait until the requests of the other participants are satisfied. This may force overloaded providers to reject requests and it is likely to cause both their revenue and reputation to be damaged. To resolve this problem, the CRBS firstly checks the exigency of requestors and then assigns resources accordingly. This scheme not only builds participants' trust in the system but also improves the resource satisfaction rate as is shown in Table 6. Moreover, it was also observed that in the case of low resource urgency a large number of transactions get confirmed in the inter-cloud scenario, which was due to the utilization of maximum budget for resource procurement, in contrast to CRBS, where agents intelligently invest to get financial benefits.

*4.1.6 Dynamic price adjustment*

To reduce lengthy negotiations, a pricing adjustment function is incorporated into the devised system which sets resource price dynamically. The pricing function needs to work fairly; if it favours a particular participant (i.e. the provider or requestor) then it is not likely to be acceptable to the other. To test the fairness of the incorporated function, typical market scenarios were considered where participants negotiate resource exchange with each other and end at a price satisfying both needs. The main objective of testing this metric was to assess the CRBS's ability to reduce lengthy negotiations by setting resources price neutrally. For evaluation purposes, the experiments were divided into three categories which were:

Category1: Requestor urgency >Provider urgency (High resource need)

Category 2: Requestor urgency <Provider urgency (Low resource need)

Category 3: Requestor urgency = =Provider urgency

In the first scenario, increasing number of requestors with high resource urgency was considered. In general, if requestors need resources urgently then they spend up to their maximum budget to get it. Likewise, if a provider determines that a requestor is eager to take resources then to gain profit the resource price is set higher. The participants negotiate with each other and, generally speaking, the negotiation concludes at a price which favours the provider. When this scenario was executed by the system the same results were achieved and the negotiation ended at a price higher than the average resource price giving more profit to resource owner as is shown in Figure 13. The x-axis in the figure represents the number of requestors involved in the experiment whereas the y-axis represents the transacting price at which the transaction was confirmed. For accuracy, the average of the transacting price for each of the number of requestors was calculated and is displayed on the graph.

The second category of experiment modeled a situation where the resource demand was substantially low and the majority of the resources were underutilized. In this scenario, to keep resources functional, a provider may opt to sell them even at a lower price. When experimented, the CRBS depicted the same behaviourand the transaction got ended at a price less than the average resource price thus giving more benefit to the requestor as is shown in Figure 14. Furthermore, it was found that the total number of transactions reduced when the resource urgency was low which was due to the agents' tendency to get the economical resources as described earlier. Similarly, in the case of equal urgencies, the transactions were made at an average price i.e. giving equal benefits to both participants as is shown in Figure 15. The experimental results validate the CRBS's ability to dynamically monitor the demand trends and to adjust the resource price appropriately i.e. giving each participant an equal chance of getting profit.

*4.2 Analysis*

A bartering platform proposed for a cloud environment should be efficient enough to handle dynamic workloads while maintaining the service availability of providers. The evaluation results revealed the same that unlike the existing inter-cloud resource exchanging platforms [5, 6, 7, 14], which works on the monetary basis, the CRBS helps providers to maintain their service provisioning without investing significantly in additional resources. Its efficient decision-making capability, as illustrated by the experiments, searches the resource pool and selects the desired resources in less than a minute. The instantaneous response helps providers to attract more customers and thus improve their business market. The negotiation mechanism of CRBS automates resource exchange by adjusting bidding price according to requestor's budget and urgency. In contrast to this, existing systems solely provide the free resources details and leave the complex processes of resource comparison and selection for requestor.

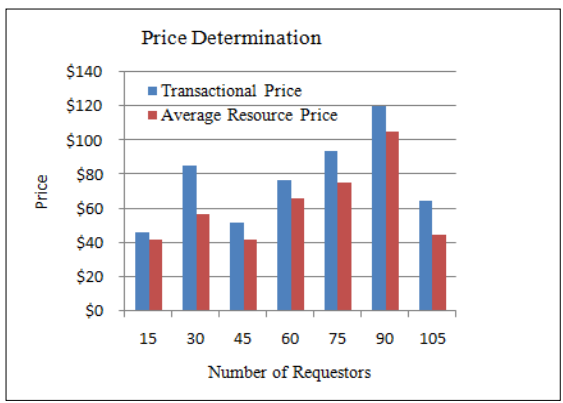

**Fig.13 Price determination in case of high resource demand**

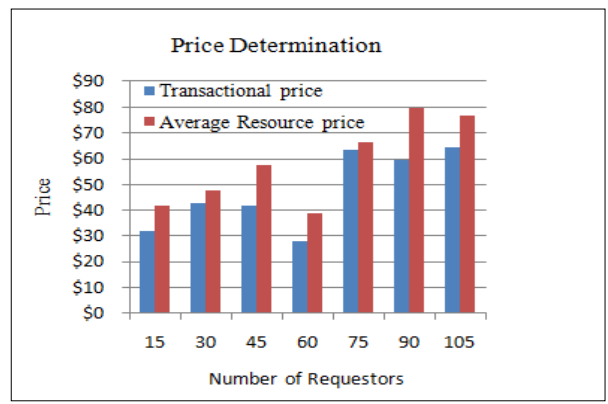

**Fig.14Price determination in case of low resource demand**

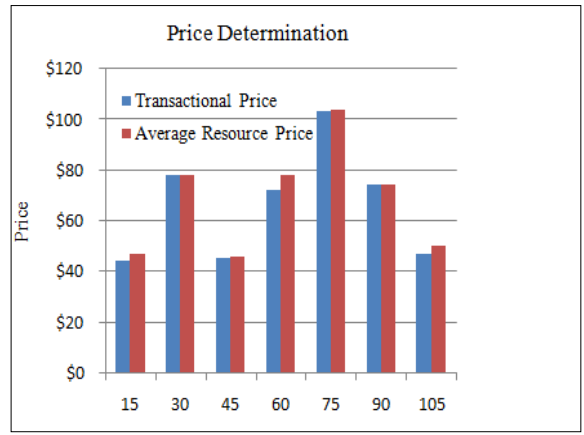

**Fig.15 Price determination in case of equal urgencies**

The urgency consideration by CRBS for resource allocation improves the overall request satisfaction rate as validated by the experiments. The current cloud platforms, however, grant resources on a first come first served strategy and may end up assigning resources to a less urgent requestor while rejecting the most urgent demands. Furthermore, the experimental results achieved by executing CRBS under varying conditions (i.e. low resource availability while having heavy resource demand, resource underutilization, and the presence of malicious users) demonstrated its responsiveness to the abrupt market situations. Under unforeseen conditions, it not only promotes fair resource exchange among participants but also adjusts resource prices dynamically i.e. heavy resource demand raises the resource price and appears beneficial for the resource owner whereas low demand results in price reduction and provides more benefits to the requestor. On the whole, the system sets resource prices impartially by providing equal opportunities for earning profit to all the participants.

The incentivizing policies of CRBS, a feature that has been neglected in the present inter-cloud systems, increased participants' tendency to remain rational throughout a transaction whereas its accounting mechanism maintains system stability by restricting participants to clear debt by sharing the same resources which were taken earlier. This feature ensures heterogeneous resources availability in resource pool and improves overall system utility. All of the discussed features corroborate CRBS suitability to work effectively under unanticipated conditions while assisting providers in maintaining consistent service delivery.

## 5    Conclusion and future work

The dynamism of the cloud paradigm is beginning to place a heavy workload on cloud providers which is difficult to manage with their primary infrastructures. To assist cloud providers in seamlessly addressing this growing workload, a new Cloud Resource Bartering System (CRBS) is presented in this work. It has been implemented by means of multi-agent technology to demonstrate the agents' applicability to managing the challenges the cloud faces. Unlike existing systems, CRBS improves the cloud services availability by enabling providers to barter resources even when they have nothing with which to reciprocate at a given time. The existence of such a platform comes together with several participation incentives such as scaling of the cloud resources to control request bursts and inter-cloud load balancing. This appears feasible for providers with limited capital and new startups that need to grow their businesses without heavily investing in their infrastructure. By managing a coalition of cloud providers, it not only saves the time to search for trustworthy bartering partners but also suggests for the most reasonable resources according to the given requirements. Furthermore, to guarantee maximum request satisfaction it allocates resources by considering requestors' urgencies. This feature enables the needy providers to timely respond to their customers and maintains their business reputations. To evaluate the effectiveness of our system, its common features have been tested against existing inter-cloud resource sharing systems. The experimental results confirm that as compared to existing resource sharing systems CRBS satisfies a large number of transactions at a given time and thus improves the overall request satisfaction and resource utilization rate. It is hoped that the outcomes of this research will prove to be a considerable advancement in the cloud domain as it will enable cloud providers to strengthen their businesses without any additional investment. To leverage the maximum benefits of CRBS, more work needs to be done on cloud standardization. The progress in this regard will resolve the interoperability issues among the various cloud platforms and ease the process of resource exchange. It will further help to improve the CRBS model by incorporating VM migration and monitoring techniques and data analysis services to forecast the likely demand for resources on a particular provider. This may enable them to reserve cloud resources in advance to accommodate surges in demands.

**Contributions**

Syeda ZerAfshan conducted the research and implemented the Resource Bartering Model under the supervision of Dr. Peter Bloodsworth. He guided her to refine research work and provided his technical expertise throughout the project. Dr. Raihan ur Rasool diligently provided valuable suggestions to improve the quality of paper throughout the process. Dr. Richard McClatchey helped during paper drafting and revision. All authors read and approved the final manuscript.